\documentclass[aps,prl,twocolumn,10pt,superscriptaddress,showpacs]{revtex4-2}
\usepackage{amsmath,amssymb,graphics,epsfig,epstopdf,color,verbatim,ulem,braket,tabularx}
\usepackage[colorlinks,linkcolor=blue,citecolor=blue,urlcolor=blue]{hyperref}
\usepackage{blindtext}
\graphicspath{{Fig_Short/}}

\begin{document}

\title{Extended Metal-Insulator Crossover with Strong Antiferromagnetic Spin Correlation in Half-Filled 3D Hubbard Model}

\author{Yu-Feng Song}
\affiliation{Institute of Modern Physics, Northwest University, Xi'an 710127, China}
\affiliation{Hefei National Laboratory for Physical Sciences at Microscale and Department of Modern Physics, University of Science and Technology of China, Hefei, Anhui 230026, China}

\author{Youjin Deng}
\email{yjdeng@ustc.edu.cn}
\affiliation{Hefei National Laboratory for Physical Sciences at Microscale and Department of Modern Physics, University of Science and Technology of China, Hefei, Anhui 230026, China}
\affiliation{Hefei National Laboratory, University of Science and Technology of China, Hefei 230088, China}

\author{Yuan-Yao He}
\email{heyuanyao@nwu.edu.cn}
\affiliation{Institute of Modern Physics, Northwest University, Xi'an 710127, China}
\affiliation{Shaanxi Key Laboratory for Theoretical Physics Frontiers, Xi'an 710127, China}
\affiliation{Hefei National Laboratory, University of Science and Technology of China, Hefei 230088, China}
\affiliation{Peng Huanwu Center for Fundamental Theory, Xian 710127, China}

\begin{abstract}
The Hubbard model at temperatures above the N\'{e}el transition, despite being a paramagnet, can exhibit rich physics due to the interplay of Fermi surface, on-site interaction $U$ and thermal fluctuations. Nevertheless, the understanding of the crossover physics remains only at a qualitative level, because of the intrinsically smooth behavior. Employing an improved variant of the {\it numerically exact} auxiliary-field quantum Monte Carlo algorithm equipped with numerical analytic continuation, we obtain a broad variety of thermodynamic and dynamical properties of the three-dimensional Hubbard model at half filling, quantitatively determine the crossover boundaries, and observe that the metal-insulator crossover regime, in which antiferromagnetic spin correlations appear strongest, exists over an extended regime in between the Fermi liquid for small $U$ and the Mott insulator for large $U$. In particular, the location of the most rapid suppression of double occupancy as $U$ increases, is found to fully reside in the metallic Fermi liquid regime, in contrast to the conventional intuition that it is a representative feature for entering the Mott insulator. Beside providing a reliable methodology for numerical study of crossover physics, our work can also serve as a timely and important guideline for the most recent optical lattice experiments.
\end{abstract}

\date{\today}
\maketitle

Mott metal-insulator transition (MIT)~\cite{Georges1996,Gebhard1997,Imada1998} has been a long-standing topic since the early days of condensed matter physics. It has been experimentally observed in various realistic materials, ranging from the typical representative of transition metal oxides~\cite{Morin1959,Carter1992,*Carter1993,Ramirez2015,Yang2011} to the latest hotspot of twisted two-dimensional (2D) moir\'{e} systems~\cite{Cao2018,Xu2020,Li2021}. While the Mott transition is mostly found to be first order~\cite{Carter1992,*Carter1993,Georges1996,Limelette2003}, recent studies show that it can also be continuous~\cite{Ghiotto2021,Tingxin2021,Kim2023}. The modern view of this phenomenon follows the original ideas of Mott~\cite{Mott1949,Mott1968} and Hubbard~\cite{Hubbard1963} that the Coulomb interaction between electrons plays a central role as splitting the conduction band and thus opening a charge gap. However, the complete understanding of Mott transition still remains a big challenge, especially in the aspect of connecting the experimental observations with Hubbard models~\cite{Ohkawa2007,Kurdestany2017,Pan2021}.

A great achievement for this problem came from the insight of dynamical mean-filed theory (DMFT) calculations in the 1990s~\cite{Georges1996,Imada1998}. In infinite dimensions where the method becomes exact, the first-order MIT with a critical end point at finite temperature as observed in experiments~\cite{Morin1959,Carter1992,*Carter1993,Ramirez2015,Yang2011} was successfully recovered in a single-band repulsive Hubbard model~\cite{Georges1992a,*Georges1992b,*Georges1993,Rozenberg1992,Zhang1993,Rozenberg1999}. Subsequent DMFT simulations for the 2D Hubbard model show similar results in paramagnetic solutions~\cite{Moukouri2001,Onoda2003,Parcollet2004,Zhang2007,Kohno2012,Ohashi2008,Park2008,Vuifmmode2013,Walsh2019,*Walsh2019L,Downey2023}. Some of these studies further suggest that, in the high temperature regime above the critical point, there might exist crossovers from the metallic state in the weakly interacting regime to the Mott insulator with strong interaction~\cite{Ohashi2008,Park2008,Vuifmmode2013,Walsh2019,*Walsh2019L,Downey2023}. Even though the system is typically a paramagnet at high temperature, the quantitative characterization and precise determination of such metal-insulator crossover (MIC) are still critical issues, due to the limited accuracy~\cite{Walsh2019,*Walsh2019L,Downey2023} and diverse signals from different observables~\cite{Svistunov2020,Kim2021}. 

As a comparison, the MIC physics in three-dimensional (3D) repulsive Hubbard model is almost in a state of emptiness. At half filling of this model, the possible Mott transition at low temperature is precluded due to the presence of the antiferromagnetic (AFM) ordered phase~\cite{Scalettar1989,Staudt2000,Werner2005,Khatami2016,Padilla2020,Iskakov2022,Lenihan2022,Kozik2013,Garioud2024,Fanjie2024}, which has been observed in a most recent optical lattice experiment~\cite{Shao2024}. Thus, a thorough study for the crossover physics in the paramagnetic (PM) phase of the half-filled 3D Hubbard model is urgently demanded to provide timely support and even guidance for more refined experimental study at intermediate to high temperatures. Besides, establishing a self-consistent and generally efficient scheme to characterize the MIC is also a very important task, and it might shed light on the MIT physics observed in realistic materials at finite temperatures. 

\begin{figure*}[ht!]
\centering
\includegraphics[width=0.90\textwidth]{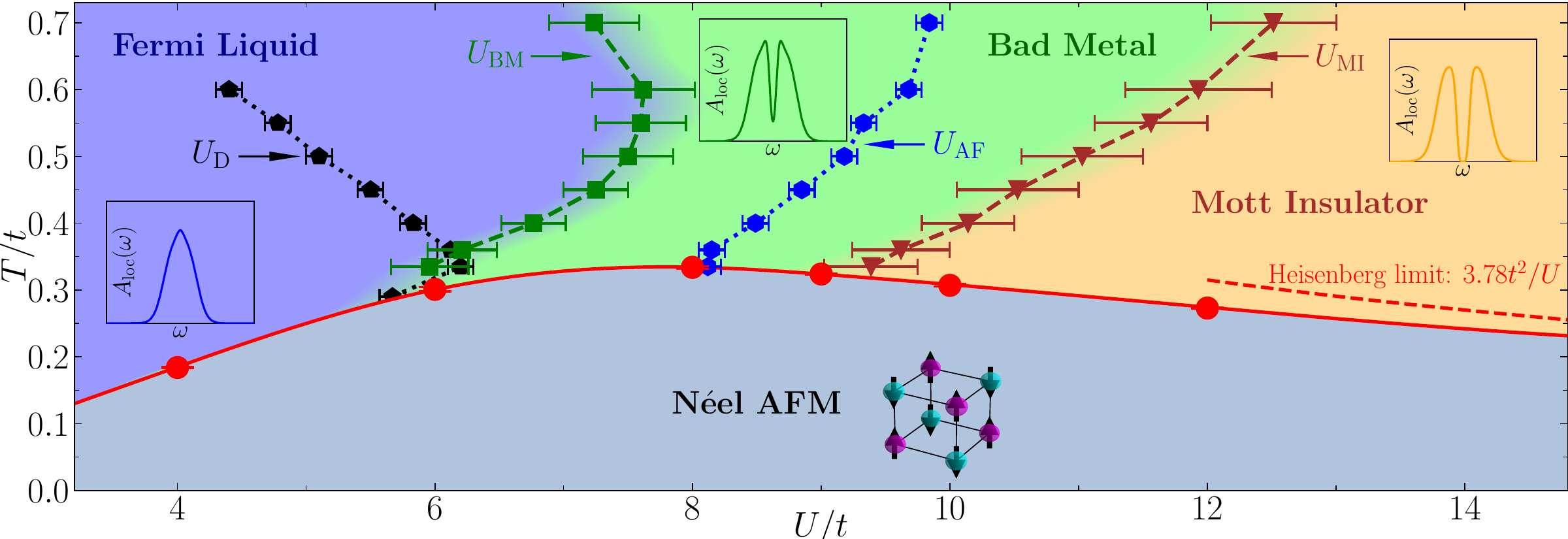}
\caption{\label{fig:Fig01PhaseDiagram} Phase diagram of half-filled 3D Hubbard model from our AFQMC calculations. Red circles show N\'{e}el transition temperatures $T_{N}$, and the solid red line connecting them is an interpolation. The dashed red line plots the result of Heisenberg limit $T_{N}=3.78t^2/U$~\cite{Sandvik1998}. Above $T_N$, Fermi liquid and Mott insulator exist in the weakly and strongly interacting regimes, respectively, and in between the bad metal emerges as the metal-insulator crossover. The onsets of bad metal as $U_{\rm BM}$ and Mott insulator as $U_{\rm MI}$ are shown by green squares and brown triangles. Crossing $U_{\rm BM}$ and $U_{\rm MI}$, the smooth crossover without any singularity is observed in all physical observables. The peak locations of AFM structure factor as $U_{\rm AF}$ are plotted as blue hexagons, which reside almost in the center of the bad metal indicating strong AFM spin correlations. The black pentagons represent the peak location of $-\partial D/\partial U$ as $U_{\rm D}$ with $D$ denoting the double occupancy. }
\end{figure*}

In this Letter, we address the above issues by systematically exploring the MIC physics above the N\'{e}el transition in the half-filled 3D Hubbard model using the auxiliary-field (AF) quantum Monte Carlo (AFQMC) method. In a numerically unbiased manner, we provide a comprehensive study for the evolution of the system along with increasing interaction strength, incorporating the numerical results of variously thermodynamic and dynamical physical observables including the single-particle spectrum, quasiparticle weight, spin-spin correlation, entropy and double occupancy. We indeed confirm the smooth crossover behaviors of the system, and obtain the boundaries of the MIC regime by carefully cross-checking different observables. Our results further show that the AFM spin correlations appear strongest in the middle of the MIC regime. Moreover, we find that the double occupancy unexpectedly fails to characterize this crossover. 

We study the half-filled 3D Hubbard model on simple cubic lattice described by $\hat{H}=\hat{H}_0+U\hat{H}_I$ with $\hat{H}_0=-t\sum_{\langle\mathbf{ij}\rangle\sigma} (c_{\mathbf{i}\sigma}^+c_{\mathbf{j}\sigma} + c_{\mathbf{j}\sigma}^+c_{\mathbf{i}\sigma})$ as the nearest-neighbor (NN) hopping and $\hat{H}_I=\sum_{\mathbf{i}}[\hat{n}_{\mathbf{i}\uparrow}\hat{n}_{\mathbf{i}\downarrow}-(\hat{n}_{\mathbf{i}\uparrow}+\hat{n}_{\mathbf{i}\downarrow})/2]$ as the on-site interaction. $\hat{n}_{\mathbf{i}\sigma}=c_{\mathbf{i}\sigma}^+ c_{\mathbf{i}\sigma}$ is the density operator with $\sigma$ ($=\uparrow$ or $\downarrow$) denoting spin. We set $t$ as the energy unit, and focus on repulsive interaction $U>0$. This model has N\'{e}el ordered ground state~\cite{Scalettar1989}, and at finite temperature it exhibits continuous AFM-PM phase transition which belongs to 3D Heisenberg university class~\cite{Campostrini2002}. We then apply the finite-temperature AFQMC algorithm~\cite{Blankenbecler1981,Hirsch1983,White1989,Scalettar1991,McDaniel2017,Yuanyao2019b,Yuanyao2019L} encoding the most recent developments~\cite{Companion} and stochastic analytic continuation (SAC) method~\cite{Sandvik2016,Shao2023} to the above model, which is sign-problem free due to the particle-hole symmetry. We perform numerical simulations for periodic supercells with $N_s=L^3$ lattice sites ($L$ as the linear system size). Our calculations reach $L=20$ for static observables and $L=12$ for dynamical properties. 

We first focus on the complete phase diagram of the above model from our AFQMC simulations as presented in Fig.~\ref{fig:Fig01PhaseDiagram}. The highest temperature in our simulations is $T/t=0.7$, much lower than the Fermi temperature $T_F/t=6$. As mentioned above, the N\'{e}el ordered phase occupies the low temperature region for all interactions. With significant improvements in precision control and dealing with finite-size effect in AFQMC calculations, we obtain highly accurate results of the N\'{e}el transition temperatures for representative interaction strengths~\cite{Companion}, via the standard finite-size scaling of AFM structure factor results up to $L=20$ using the known critical exponents~\cite{Campostrini2002}. Our results illustrate the highest transition temperature $T_N$$\sim$$0.33t$ achieved around $U/t=8$. These are comparable to the results in previously unbiased calculations~\cite{Staudt2000,Fanjie2024,Kozik2013,Garioud2024}. 

The contents above the N\'{e}el transition in the phase diagram summarize the main results of this work. In the weakly interacting regime, the correlated Fermi liquid state appears as an adiabatical evolution from $U=0$, and its fingerprint signature is the coherence peak in local single-particle spectral function $A_{\rm loc}(\omega)$ at Fermi energy. Approaching the large $U$ limit, the system should be a Mott insulator exihiting a fermionic gap as $A_{\rm loc}(\omega\sim0)=0$, which survives as long as the temperature energy scale $\sim$$k_BT$ is smaller than the $T=0$ gap of the system. Between these two limits, our numerical results clearly reveal an intermediate regime for which the spectra $A_{\rm loc}(\omega)$ around Fermi energy shows a dip. We adopt the term ``bad metal''~\cite{Deng2013,Ding2019,BadMetalNote} for this MIC regime. The three regimes with their $A_{\rm loc}(\omega)$ features (as insets) and the boundaries defined as the onsets of bad metal (as $U_{\rm BM}$) and Mott Insulator (as $U_{\rm MI}$) from our simulations are shown in Fig.~\ref{fig:Fig01PhaseDiagram}. Additional results of peak locations of AFM structure factor as $U_{\rm AF}$ and the most rapid suppression of double occupancy as $U_{\rm D}$ versus $U/t$ are also included in the plot. The former resides around the center of the bad metal regime meaning that this crossover region has strong AFM spin correlations. The latter steps into the Fermi liquid regime indicating that the double occupancy has little connection with the MIC in this system within $T/t>0.36$. These signatures are obtained with $L=12$ for $T/t\le0.36$ and $L=8$ for higher temperatures (as reported in this work), which are verified to have negligible finite-size effect.

\begin{figure}[t]
\centering
\includegraphics[width=0.99\columnwidth]{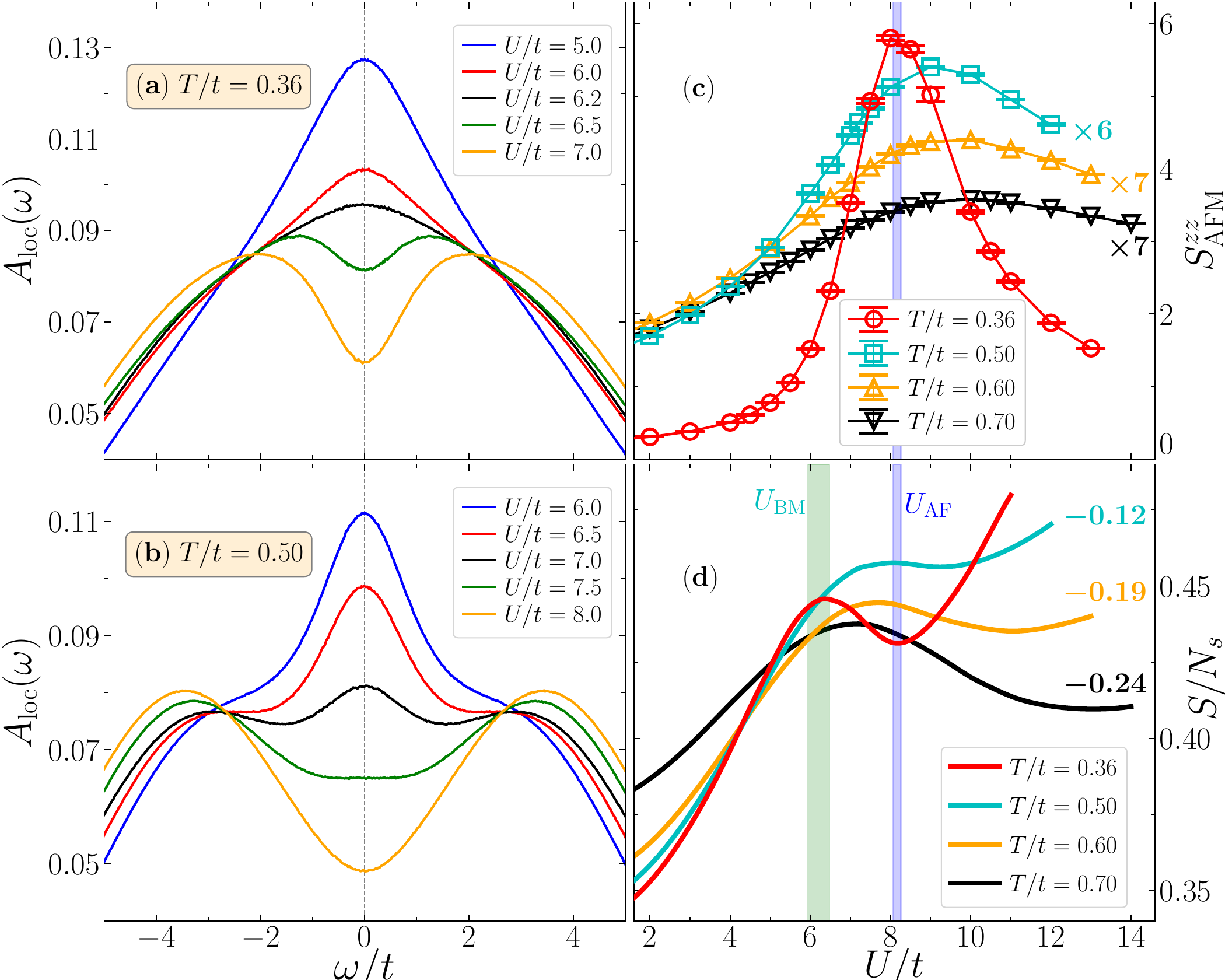}
\caption{\label{fig:Fig02Spectrum} Local single-particle spectral function $A_{\rm loc}(\omega)$ crossing the onset of bad metal ($U_{\rm BM}$) for (a) $T/t=0.36$ and (b) $T/t=0.50$. Panels (c) and (d) plot the AFM structure factor $S^{zz}_{\rm AFM}$ and thermal entropy per site $S/N_s$ for four temperatures, respectively. For $T/t=0.50,0.60,0.70$, the $S/N_s$ results are shifted by $-0.12,-0.19,-0.24$, and $S^{zz}_{\rm AFM}$ data are rescaled by factors $\times6,\times7,\times7$, to fit into the plots. The uncertainty of $S/N_s$ is indicated by the thickness of the lines. For $T/t=0.36$, the result of $U_{\rm BM}$ determined from $A_{\rm loc}(\omega)$ in (a) as the green shading and the peak location range of $S^{zz}_{\rm AFM}$ as the blue shading are shown in (c) and (d). The system is $L=12$ for $T/t=0.36$ and $L=8$ for other temperatures.}
\end{figure}

We first probe the onset of bad metal $U_{\rm BM}$ via the local spectral function $A_{\rm loc}(\omega)$, which is obtained from the dynamic single-particle Green's function $G_{\rm loc}(\tau)=(2N_s)^{-1}\sum_{\mathbf{i}\sigma}\langle c_{\mathbf{i}\sigma}(\tau)c_{\mathbf{i}\sigma}^+\rangle$ via the SAC method~\cite{Sandvik2016,Shao2023}. The $A_{\rm loc}(\omega)$ results for $T/t=0.36$ and $T/t=0.50$ are plotted in Figs.~\ref{fig:Fig02Spectrum}(a) and \ref{fig:Fig02Spectrum}(b) [Note that $A_{\rm loc}(\omega)$ is symmetric about $\omega=0$ at half filling]. For increasing $U/t$, we see that $A_{\rm loc}(\omega)$ around $\omega=0$ is gradually suppressed and it smoothly evolves from the coherence peak structure to the shape of two broaden peaks at finite $\omega$ indicating the formation of upper and lower Hubbard bands. This dip signature emerges due to the lack of low-energy fermionic excitations, and thus marks the entrance into the bad metal regime. We define $U_{\rm BM}$ as the interaction strength signifying the disappearance of the coherence peak of $A_{\rm loc}(\omega)$, and numerically determine it via extrapolating $Q=|A_m-A_{\rm loc}(\omega=0)|/A_m$ to zero versus $U$ (with $A_m$ as the local minimum or maximum of $A_{\rm loc}(\omega)$~\cite{Companion}). And its uncertainty is estimated by considering a $10\%$ drop of $A_{\rm loc}(\omega=0)$ comparing to the value at the left and right fringes~\cite{Bauer2014}, to account for the possible uncertainty in SAC calculations. Such procedure acquires the results $U_{\rm BM}=6.20(30)$ for $T/t=0.36$ and $U_{\rm BM}=7.50(35)$ for $T/t=0.50$.

Besides the dip of $A_{\rm loc}(\omega)$, the system also shows intriguing behaviors regarding the spin fluctuation when crossing $U_{\rm BM}$ and entering the bad metal regime. We quantify this property via the AFM structure factor defined as $S_{\rm AFM}^{zz} = N_s^{-1}\sum_{\mathbf{ij}}(-1)^{\mathbf{i}+\mathbf{j}}\langle \hat{s}_{\mathbf{i}}^z \hat{s}_{\mathbf{j}}^z \rangle$, and present its results in Fig.~\ref{fig:Fig02Spectrum}(c). We observe that $S_{\rm AFM}^{zz}$ increases rapidly around $U_{\rm BM}$, reach the maximum, and then drops towards the strong interaction limit. The peak and the successive decrease of $S_{\rm AFM}^{zz}$ with increasing $U/t$ can be understood from reducing the coupling constant $J=4t^2/U$ of the effective Heisenberg model description of the system. We obtain the peak location of $S_{\rm AFM}^{zz}$ as $U_{\rm AF}$, and see that it almost follows the center of the bad metal regime as shown in Fig.~\ref{fig:Fig01PhaseDiagram}. Similar results can be obtained for the NN spin-spin correlation and spin correlation length with comparable values of $U_{\rm AF}$~\cite{Companion}. These together manifest the strong AFM spin correlations as an accompanying property of the bad metal. 

We further find that the thermal entropy $S$ can reproduce the above results of $U_{\rm BM}$, and it can serve as a bridge to connect the behaviors of $A_{\rm loc}(\omega)$ and $S_{\rm AFM}^{zz}$ in the crossover regime. Here we have developed an efficient scheme to calculate $S$ versus $U$ at fixed temperature $T$ via $S(U)=T^{-1}\big(\langle\hat{H}\rangle-F_0-\int_0^{U}\langle\hat{H}_I\rangle dU^{\prime}\big)$~\cite{Companion} with $F_0$ as $U=0$ free energy. This scheme only involves AFQMC simulations at that fixed temperature and thus substantially reduces the computational effort compared to the conventional method via the integral of total energy~\cite{Padilla2020}. In Fig.~\ref{fig:Fig02Spectrum}(d), we show the entropy per site $S/N_s$ for the same temperatures as Fig.~\ref{fig:Fig02Spectrum}(c). With increasing $U/t$, the entropy first increases and reaches a maximum. We verify that the peak location is well consistent with $U_{\rm BM}$ for all the temperatures. The reason for this coincidence is that, in correlated Fermi liquid, the entropy is proportional to the effective mass of fermions~\cite{Walsh2019,*Walsh2019L,Downey2023} which generally grows with increasing interaction. Thus, the peak of entropy can be taken as the termination of the Fermi liquid regime and the entrance into the crossover. For $U>U_{\rm BM}$, the entropy decreases and develops local minimum. We find that its location is well consistent with $U_{\rm AF}$ (peak location of $S_{\rm AFM}^{zz}$) for $T/t\le0.50$ while it is larger than $U_{\rm AF}$ for higher temperatures. This could be explained by decomposing contributions to the entropy into charge and spin channels for $U>U_{\rm BM}$, i.e., $S=S_{\rm c}+S_{\rm s}$. For charge channel, $S_{\rm c}$ should track the result of $A_{\rm loc}(\omega=0)$, while $S_{\rm s}$ from spin channel behaves oppositely with $S_{\rm AFM}^{zz}$. Thus, once entering bad metal regime, $S_{\rm c}$ should monotonically decrease while $S_{\rm s}$ first decreases to a local minimum and then increases again and finally saturates to $\ln 2$ reaching $U=\infty$ limit. This also highlights the peak location of entropy as the onset of bad metal. Moreover, for $T/t\le0.50$, $A_{\rm loc}(\omega=0)$ at $U=U_{\rm AF}$ is quite small, indicating vanishing $S_{\rm c}$. As a result, the local minimum of the entropy is dominated by $S_{\rm s}$, and thus its location conforms with $U_{\rm AF}$. However, for higher temperatures, the contribution of $S_{\rm c}$ becomes more significant due to more charge excitations. This together with the valley structure of $S_{\rm s}$ renders the local minimum of total entropy to appear at slightly larger $U$ than $U_{\rm AF}$, consistent with the results of $T/t=0.60$ and $0.70$ as shown in Fig.~\ref{fig:Fig02Spectrum}(d).

\begin{figure}
\centering
\includegraphics[width=0.90\columnwidth]{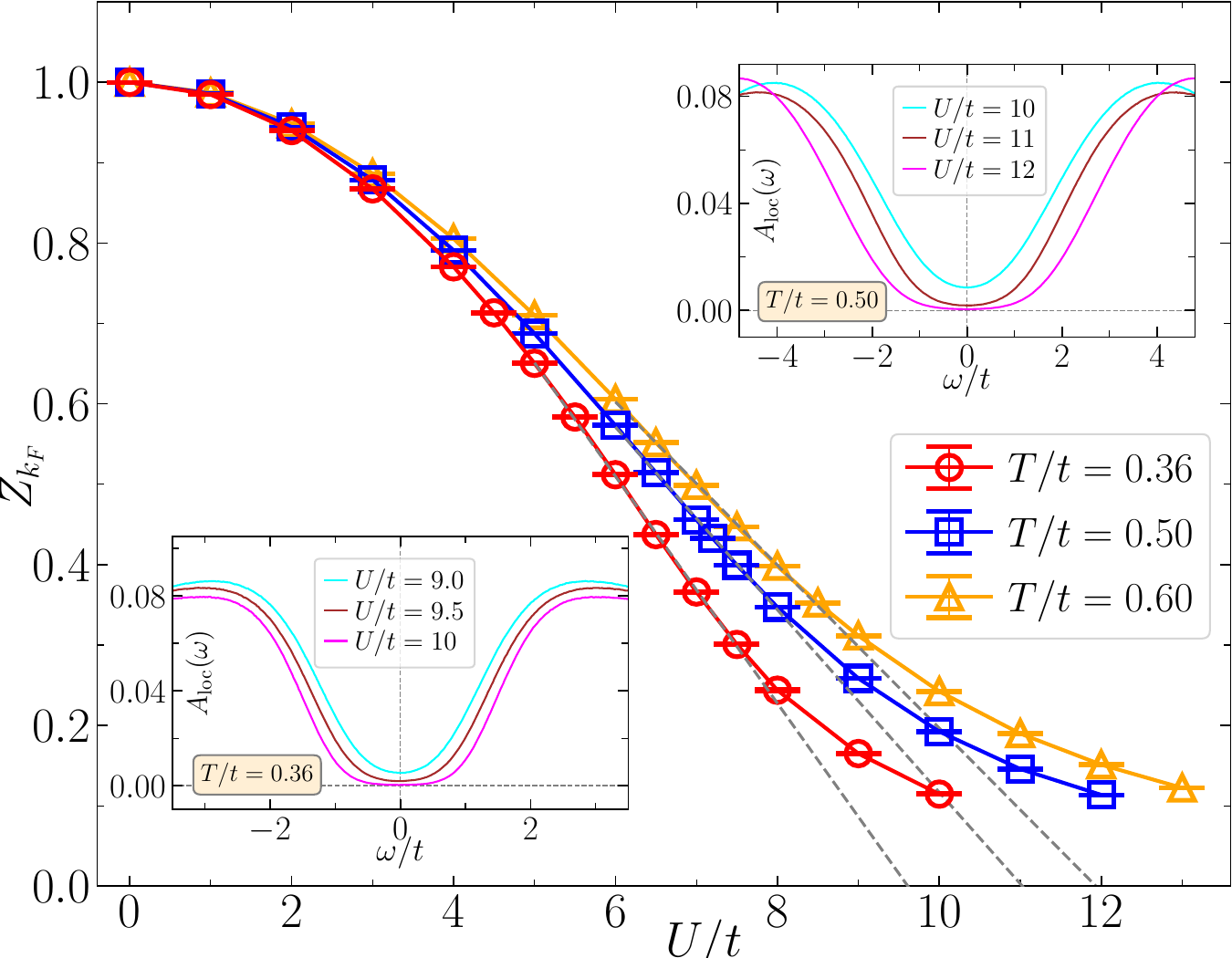}
\caption{\label{fig:Fig03QWeight} Determination of the onset of Mott insulator ($U_{\rm MI}$). The main plot shows the quasiparticle weight $Z_{k_F}$ versus $U/t$ for $T/t=0.36,0.50,0.60$. Linear fittings (gray dashed lines) are performed for the AFQMC data in the middle range, and the intercepts of $U/t$ corresponding to extrapolated $Z_{k_F}=0$ are taken as $U_{\rm MI}$. The local single-particle spectral function $A_{\rm loc}(\omega)$ for $T/t=0.36$ and $0.50$ plotted in the insets confirm that $U_{\rm MI}$ obtained from $Z_{k_F}$ indeed corresponds to the entrance into Mott insulator with $A_{\rm loc}(\omega=0)\sim0$. The system size is the same as Fig.~\ref{fig:Fig02Spectrum}. }
\end{figure}

We then turn to the onset of Mott insulator $U_{\rm MI}$ in the phase diagram. To avoid the ambiguity of $A_{\rm loc}(\omega=0)=0$, we first determine $U_{\rm MI}$ from the quasiparticle weight at the Fermi surface computed by $Z_{k_F}\approx[1-\text{Im}\Sigma_{\sigma}(\mathbf{k}_F,i\omega_0)/\omega_0]^{-1}$~\cite{Liebsch2003}, and then combine the results of $A_{\rm loc}(\omega=0)$ as a supplementary tool to estimate the uncertainty. We compute the self energy $\Sigma_{\sigma}(\mathbf{k}_F,i\omega_0)$ with $\mathbf{k}_F$ as Fermi wave vector and $\omega_0=\pi/\beta$ via the Dyson equation. We have performed additional average of $Z_{k_F}$ results at all $\mathbf{k}_F$ points since we find that the system is highly isotropic at the Fermi surface for various properties. The number $(1-Z_{k_F})$ measures the interaction induced transfer of spectral weight around $\omega=0$ to the incoherent Hubbard bands. Thus the exact $Z_{k_F}=0$, which only exists at $T=0$ for finite $U$, marks the Mott insulator phase. The results of $Z_{k_F}$ versus $U/t$ for $T/t=0.36$, $0.50$, and $0.60$ are shown in the main plot of Fig.~\ref{fig:Fig03QWeight}. With increasing $U/t$, the quantity decreases from unity in the noninteracting limit to a small residual value rounded off by the finite temperature. We take $U_{\rm MI}$ as the interaction strength where a linear extrapolation of $Z_{k_F}$ with intermediate $U/t$ reaches zero (gray dashed lines in Fig.~\ref{fig:Fig03QWeight})~\cite{Liebsch2003}. We estimate its uncertainty by incorporating the criteria $A_{\rm loc}(\omega=0)<\epsilon$ (with $\epsilon\sim 10^{-3}$ as a threshold). The corresponding results of $A_{\rm loc}(\omega)$ around $U_{\rm MI}$ are shown in insets of Fig.~\ref{fig:Fig03QWeight}. This generates $U_{\rm MI}=9.6(4)$ for $T/t=0.36$ and $U_{\rm MI}=11.0(5)$ for $T/t=0.50$. We also find that the charge compressibility $\chi_e$ becomes tiny around $U_{\rm MI}$ and similar criteria $\chi_e<\epsilon$ presents well consistent results for $U_{\rm MI}$~\cite{Companion}. Combining results at different temperatures, the full curve of $U_{\rm MI}$ in Fig.~\ref{fig:Fig01PhaseDiagram} shows almost linear dependence with temperature. This reveals the nature of the Mott insulator regime: the $T=0$ gap of the system which is proportional to $U$ overtakes the temperature energy scale $\sim$$k_BT$.

\begin{figure}
\centering
\includegraphics[width=0.90\columnwidth]{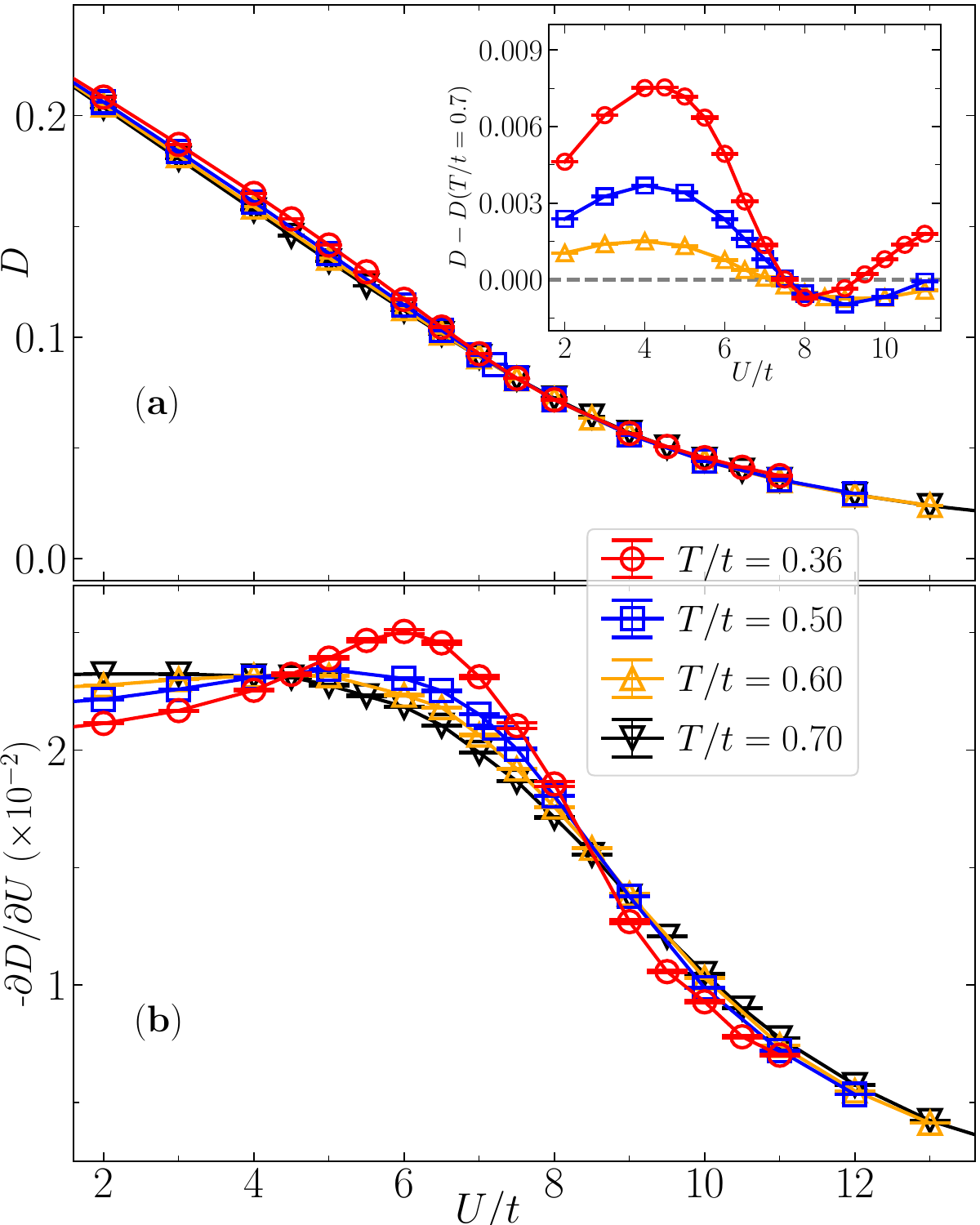}
\caption{\label{fig:Fig04DouOcc} Double occupancy $D$ and its first-order derivative $-\partial D/\partial U$ versus $U/t$. The inset in (a) shows the relative $D$ differences with that of $T/t=0.70$. In (b), the peak location of $-\partial D/\partial U$ is taken as $U_{\rm D}$ for $T/t\le0.60$, while there is no peak for $T/t=0.70$. Note that $-\partial D/\partial U$ results in (b) are directly computed from AFQMC simulations instead of the numerical differentiation. The system size is same as Fig.~\ref{fig:Fig02Spectrum}. }
\end{figure}

The double occupancy $D=N_s^{-1}\sum_{\mathbf{i}}\langle\hat{n}_{\mathbf{i}\uparrow}\hat{n}_{\mathbf{i}\downarrow}\rangle$ is usually applied as a tool to detect the Mott MIT~\cite{Rozenberg1999,Liebsch2003,Parcollet2004,Arita2005,Hoshino2017,Zhu2019} in Hubbard models, based on its relation with total energy $E_g$ as $D=\partial E_g/\partial U$ at $T=0$ and with the free energy $F$ as $D=\partial F/\partial U$ at finite temperature. It typically exhibits discontinuity and an inflection point around the first-order and continuous MIT, respectively. In the system we study, the transition is replaced by MIC, and thus it represents a fully new situation. We present the numerical results of $D$ and $-\partial D/\partial U$ versus $U/t$ in Fig.~\ref{fig:Fig04DouOcc}. Note that, instead of numerical differentiation, we compute the derivative using $-\partial D/\partial U=2N_s^{-1}\int_{0}^{\beta/2}C_{\hat{H}_I}(\tau,0)d\tau$~\cite{Companion} with $C_{\hat{H}_I}(\tau, 0)=\langle\hat{H}_I(\tau)\hat{H}_I(0)\rangle - \langle\hat{H}_I(\tau)\rangle\langle\hat{H}_I(0)\rangle$ directly measured in AFQMC simulations. We see that $D$ simply decreases with increasing $U/t$, and its temperature variation is quite small for $0.36\le T/t\le 0.70$. In the inset of Fig.~\ref{fig:Fig04DouOcc}(a), we also present the differences of $D$ relative to $T/t=0.70$. For all displayed $U/t$ except a small region around $U/t=8$, a clear decrease of $D$ upon heating can be observed in a specific and $U$-dependent range of $T$ [as $(0.335,0.80)$ for $U/t=6$ and $(0,0.54)$ for $U/t=10$], which resembles the Pomeranchuk effect in $^3$He~\cite{Richardson1997} and was also observed in Refs.~\cite{Khatami2016,Padilla2020}. This phenomenon can be understood from the entropy results in Fig.~\ref{fig:Fig02Spectrum}(d) via Maxwell's relation $-(\partial D/\partial T)_{U}=(\partial S/\partial U)_{T}$, rendering the increase of $S$ with enhancing $U$ equivalent to the decrease of $D$ with raising temperature. Moreover, for $T/t\le 0.60$, the $D$ curve has inflection point as signified by the broden peak in $-\partial D/\partial U$ results in Fig.~\ref{fig:Fig04DouOcc}(b), while such behaviors disappear for $T/t\ge 0.70$. Collecting the peak locations of $-\partial D/\partial U$ for different temperatures forms the $U_{\rm D}$ curve in Fig.~\ref{fig:Fig01PhaseDiagram}, which fully resides in the Fermi liquid regime and even moves towards smaller $U$ for $T/t > 0.36$. These unbiased results reveal that the double occupancy fails to capture the MIC physics. We have also calculated other quantities including charge compressibility, fidelity susceptibility and Matsubara-frequency self-energy, and find that they share similar behavior as the double occupancy regarding the MIC in the PM phase~\cite{Companion}. 

In summary, we have revealed that, in the half-filled 3D Hubbard model above the N\'{e}el transition, a metal-insulator crossover regime exists between the Fermi liquid and Mott insulator in a significant range of interaction strength. And this crossover regime is accompanied by strong AFM spin correlations. We have established an efficient scheme to characterize the crossover and compute its boundaries combining various observables. The onset of the crossover regime can be determined from single-particle spectrum and thermal entropy, while its termination is manifested by the spectrum, quasiparticle weight and charge compressibility. This scheme is very likely to contribute to similar studies for 2D systems~\cite{Walsh2019,*Walsh2019L,Downey2023,Svistunov2020,Kim2021}, with frustrations~\cite{Parcollet2004} and even for realistic materials~\cite{Limelette2003}. We have also found that some commonly used observables are not able to characterize the crossover physics, especially the double occupancy. Our work belongs to the only few unbiased studies for the metal-insulator crossover physics in lattice models. On the other hand, considering the experimental capability of accessing the AFM ordered phase~\cite{Shao2024}, our results can provide timely and important benchmarks and even guidelines for the next-step uniform optical lattice experiment of large scale in the intermediate to high temperature regime. 

\begin{acknowledgments}
Y.-Y. H. acknowledges Yang Qi, Gang Li, Mingpu Qin, Hui Shao and Xiao Yan Xu for valuable discussions. This work was supported by the National Natural Science Foundation of China (under Grants No. 12247103, No. 12204377, and No. 12275263), the Innovation Program for Quantum Science and Technology (under Grant No. 2021ZD0301900), the Natural Science Foundation of Fujian province of China (under Grant No. 2023J02032), and the Youth Innovation Team of Shaanxi Universities.
\end{acknowledgments}

\bibliography{3DHubbardRef_Short}

\begin{thebibliography}{72}
\bibitem [{Bad()}]{BadMetalNote}%
  \BibitemOpen
  \href@noop {} {}\bibinfo {note} {We note that some papers use the term ``Bad
  Metal'' for the regime where the system exhibits linear resistivity versus
  temperature. This typically happens in higher temperature regime as
  $T/t>1.0$. Here, we adopt ``Bad Metal'' only to describe the metal-insulator
  crossover regime in which $A_{\rm loc}(\omega)$ has a dip around $\omega=0$
  but with $A_{\rm loc}(\omega=0)>0$. This Bad Metal regime can be also
  regarded as an unusual Fermi liquid state at intermediate to high
  temperatures, in the sense that the quasiparticle weight $Z$ is finite and
  changes smoothly.}\BibitemShut {Stop}%
\end{thebibliography}

\end{document}